\documentclass[11pt,twoside]{article}
\usepackage{asp2014}

\aspSuppressVolSlug
\resetcounters

\begin{document}

\title{Science with an ngVLA: ngVLA Observations of Coronal Magnetic Fields}
\author{G. D. Fleishman$^1$, G. M. Nita$^1$, S. M. White$^2$, D. E. Gary$^1$, T. S. Bastian$^3$\\
\affil{$^1$Center For Solar-Terrestrial Research, New Jersey Institute of Technology, Newark, NJ 07102, USA; \email{gfleishm@njit.edu}}
\affil{$^2$Space Vehicles Directorate, Air Force Research Laboratory, Kirtland AFB, NM 87123, USA; \email{stephen.white.24@us.af.mil}}
\affil{$^3$National Radio Astronomy Observatory, Charlottesville, VA, USA; \email{tbastian@nrao.edu}}}

\paperauthor{G. D. Fleishman}{gfleishm@njit.edu}{0000-0001-5557-2100}{New Jersey Institute of Technology}{Center For Solar-Terrestrial Research}{Newark}{NJ}{07102}{USA}
\paperauthor{G. M. Nita}{gelu.m.nita@njit.edu}{}{New Jersey Institute of Technology}{Center For Solar-Terrestrial Research}{Newark}{NJ}{07102}{USA}
\paperauthor{S. M. White}{stephen.white@us.af.mil}{}{Space Vehicles Directorate}{Air Force Research Laboratory}{Kirtland AFB}{NM}{87123}{USA}
\paperauthor{D. E. Gary}{dale.gary@njit.edu}{}{New Jersey Institute of Technology}{Center For Solar-Terrestrial Research}{Newark}{NJ}{07102}{USA}
\paperauthor{T. S. Bastian}{tbastian@nrao.edu}{ORCID_Or_Blank}{National Radio Astronomy Observatory}{}{Charlottesville}{VA}{22903}{USA}

\begin{abstract}
Energy stored in the magnetic field in the solar atmosphere above active regions is a key driver of all solar activity (e.g., solar flares and coronal mass ejections), some of which can affect life on Earth. Radio observations provide a unique diagnostic of the coronal magnetic fields that make them a critical tool for the study of these phenomena, using the technique of broadband radio imaging spectropolarimetry. Observations with the ngVLA will provide unique observations of coronal magnetic fields and their evolution, key inputs and constraints for MHD numerical models of the solar atmosphere and eruptive processes, and a key link between lower layers of the solar atmosphere and the heliosphere. In doing so they will also provide practical "research to operations" guidance for space weather forecasting.
\end{abstract}

\section{Introduction}

Magnetic fields are important in a variety of astrophysical contexts. For example, the stored magnetic energy in an extended volume of magnetized plasma can be rapidly, catastrophically released, often accompanied by changes in the magnetic connectivity. The corresponding driving process has come to be known as magnetic reconnection, and understanding this phenomenon is central to many problems of astrophysics, such as the origin of the radio arc in the galactic center \citep{2005PASJ...57L..39S}, quasar superluminal sources \citep{2005A&A...441..845D}, the hot component of the galactic ridge X-ray emission \citep{1999PASJ...51..161T}, particle acceleration in the accretion disks of AGNs \citep{1998A&A...335...26S}, and solar-analogous problems in stellar astrophysics, such as how stellar atmospheres are heated and nonthermal particles are produced in active, highly magnetized stars \citep{2005Natur.434.1098H}. Thanks to its proximity and the availability of many observing facilities across a wide range of wavelengths, the Sun is arguably the best astrophysical laboratory for obtaining clues to magnetic field structure, its evolution, and magnetic energy release.

A key area for such study is coronal magnetic fields in solar active regions. Active regions are the locations on the Sun where strong magnetic fields generated deep in the solar convection zone rise through the photosphere, often in the guise of sunspots, up into the corona. Active region magnetic field evolution and dynamics within the corona then drive many of the energetic phenomena that we observe on the Sun, converting magnetic energy into other forms.

While currently available space- and ground-based solar optical telescopes are already capable of precise measurements of the photospheric magnetic field with sub-arcsecond angular resolution and high temporal resolution using the Zeeman effect, these techniques require appropriate spectral lines and great sensitivity to be used in the corona. Next generation instruments like Daniel K. Inouye Solar Telescope \citep[DKIST;][]{2016AN....337.1064T} will exploit such measurements using near-infrared coronal spectral lines \citep{2018ApJ...852...52D}, but these will only be effective above the solar limb. As we show here, radio observations are positioned to be powerful and versatile diagnostic of coronal magnetic fields using the ngVLA.

The absence of quantitative measurements of coronal magnetic fields to date has led the community to rely heavily on magnetic field extrapolations that use measurements at the photosphere as the boundary condition. For example, modern nonlinear force-free field (NLFFF) extrapolation algorithms can be used to produce models of the coronal field \citep[e.g.,][]{2009ApJ...696.1780D, 2015ApJ...811..107D}, but owing to the fact that the photosphere is not actually force-free, as well as other observational and theoretical limitations, such extrapolations are far from unique \citep{2015ApJ...811..107D}. Therefore, the coronal magnetic field modeling based on such extrapolations requires modification and independent verification using observational diagnostics of the magnetic field at coronal heights. Below we briefly review the available radio techniques for probing the coronal magnetic and thermal structures, and highlight synergies between the radio diagnostics and other data sets and models.

\begin{figure}\centering
\includegraphics[width=1\textwidth, clip]{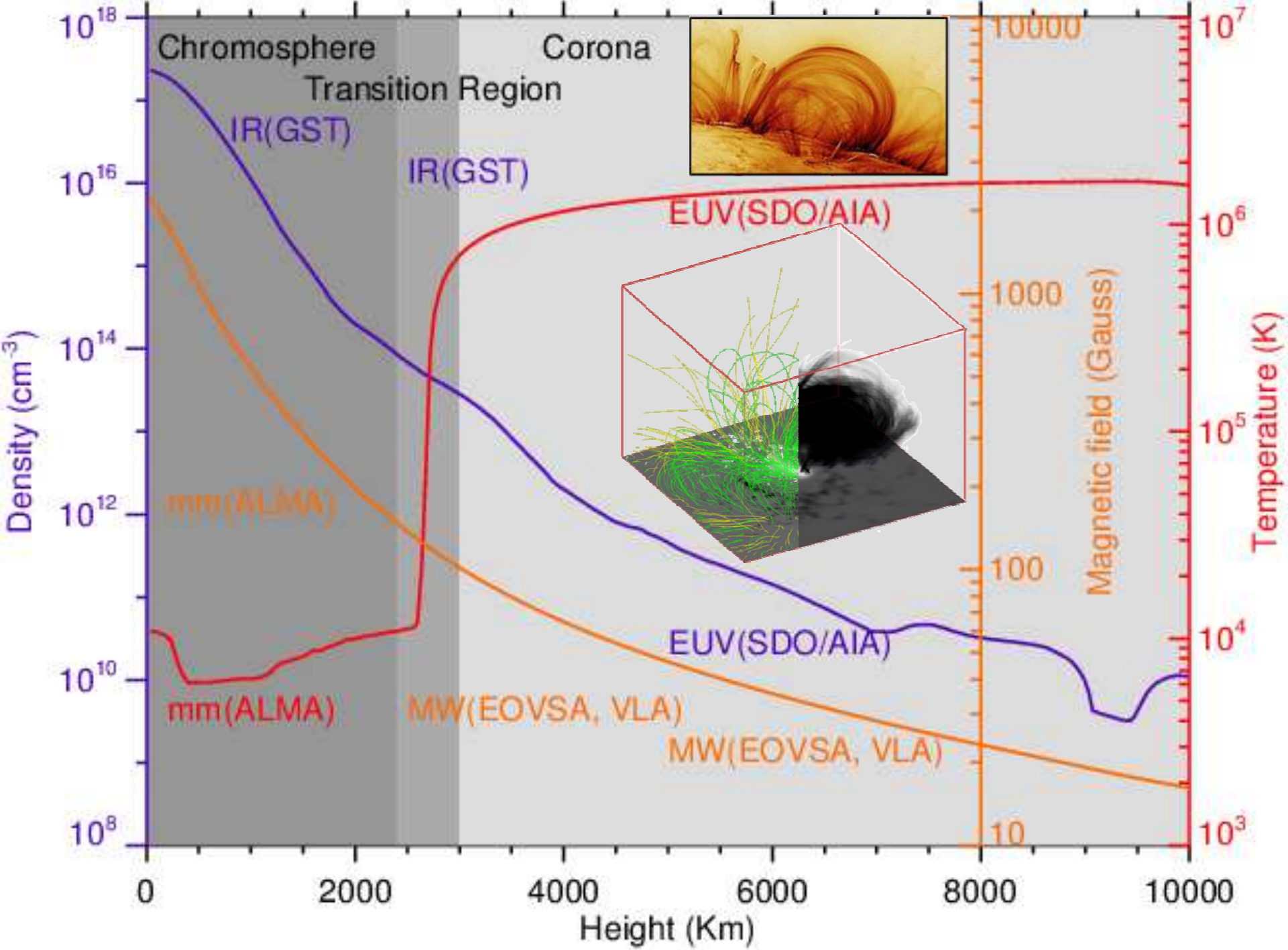} 
\caption{\label{f_intro} Schematic 1D structure of the solar atmosphere taken from a single line of sight from the MHD Bifrost model \citep{2016A&A...585A...4C}: total density (blue), temperature (red), and magnetic field (orange). The dense chromosphere and tenuous corona are separated by the relatively thin transition region-layer in which the temperature sharply rises, from thousands to millions of Kelvin. The physical parameters that can be diagnosed in each region by several ground-based and space-borne instruments, and their observational windows, are indicated in corresponding colors: GST and DKIST: Infrared (IR); EOVSA and the VLA: dm/cm wavelengths; ALMA: mm/submm wavelenghts; SDO/AIA: EUV. Insets illustrate a 2D SDO/AIA image of coronal loops, and a 3D coronal model visualized with a set of magnetic field lines (left) and electron density distribution (right).
 }
\end{figure}

\section{Sensitivity of Radio Emission to Coronal Magnetic Fields}

Radio emission processes are well known to depend strongly on the magnetic field strength in the source region due to two effects: (1) the ${\bf v}\times{\bf B}$ Lorentz force causes the emitting electrons to spiral in the magnetic field to produce gyroemission, which takes the form of gyroresonance (GR) emission for thermal electrons in the hot corona of active regions, and gyrosynchrotron (GS) emission for mildly-relativistic electrons in flares; and (2) the plasma itself is birefringent, with differing indices of refraction \citep[e.g.,][]{Fl_Topt_2013_CED} for the ordinary (o) and extraordinary (x) magnetoionic modes, which leads to a magnetic-field-dependent polarization of each type of emission, including free-free (FF) bremsstrahlung emission \citep[e.g.,][]{2004ASSL..314..115G}.  Hence, both Stokes I and V polarization parameters must be observed. \footnote{The Faraday depth of the corona is so large that Stokes Q and U are unobservable except, possibly, under special circumstances. See Allisandrakis \& Chiuderi-Drago 1995.}

All three emission mechanisms, GR, GS, and FF, have been exploited in studies of radio emission from the Sun to estimate magnetic field strengths in the corona.  However, such estimates based on either spatial or spectral information alone generally suffer from a large number of ambiguities that require varying degrees of assumptions to be made, or else require special conditions where the assumptions and ambiguities can be minimized.  These ambiguities can be largely eliminated when both spatial and spectral information in both senses of circular polarization are available simultaneously. In this case, three-dimensional data cubes are used to provide spatially-resolved, polarized brightness-temperature spectra over an appropriate (broad) frequency range. We refer to this as {\sl microwave imaging spectropolarimetry}. It is precisely this capability that ngVLA will provide, permitting the full exploitation of the magnetic field diagnostics of GR, GS, and FF emission over an unprecedentedly broad spectral band.

For active regions, the two emission mechanisms relevant for the relatively cool chromosphere and hot, magnetized corona are the GR and FF mechanisms, both of which are well understood. As described by \citet{2004ASSL..314...71G}, spatially-resolved brightness temperature spectra $T_b(f)$, which can be obtained from any pixel of the multi-frequency image data cube, provide temperature as a function of frequency, $T(f)=T_b(f)$, provided the radio emission is optically thick. The optical depth of FF emission varies as $n_e^2\,f^{-2}\,T^{-1.5}$, where $n_e$ is the ambient electron density, so it favors dense cool regions and low frequencies. The corona is seldom optically thick to free-free absorption at frequencies $\gtrsim$3 GHz, at which point optically thick GR emission becomes relevant if the coronal magnetic field is sufficiently strong.  Typically, GR emission has a significant optical depth at the lowest few harmonics $s = $1, 2, 3 (more seldom 4 or 5) of the gyrofrequency $f_B$, $f = sf_B = 2.8 sB$ MHz, which allows the conversion $T(f) \Rightarrow T(sB)$; \citet{2011ApJ...728....1T,Nita_etal_2011,2015ApJ...805...93W}. Given that the temperature of the solar atmosphere drops precipitously in a narrow height range at the transition region (Fig.~\ref{f_intro}), the base of the corona, the $T_b$ spectrum likewise cuts off at those frequencies whose GR emission originates in the transition region. It is straightforward to then deduce the magnetic field strength at the base of the corona \citep[e.g.,][]{2007SSRv..133...73L} to produce so-called level-0 coronal magnetic field maps.

\begin{figure}\centering
\includegraphics[width=0.95\textwidth, clip]{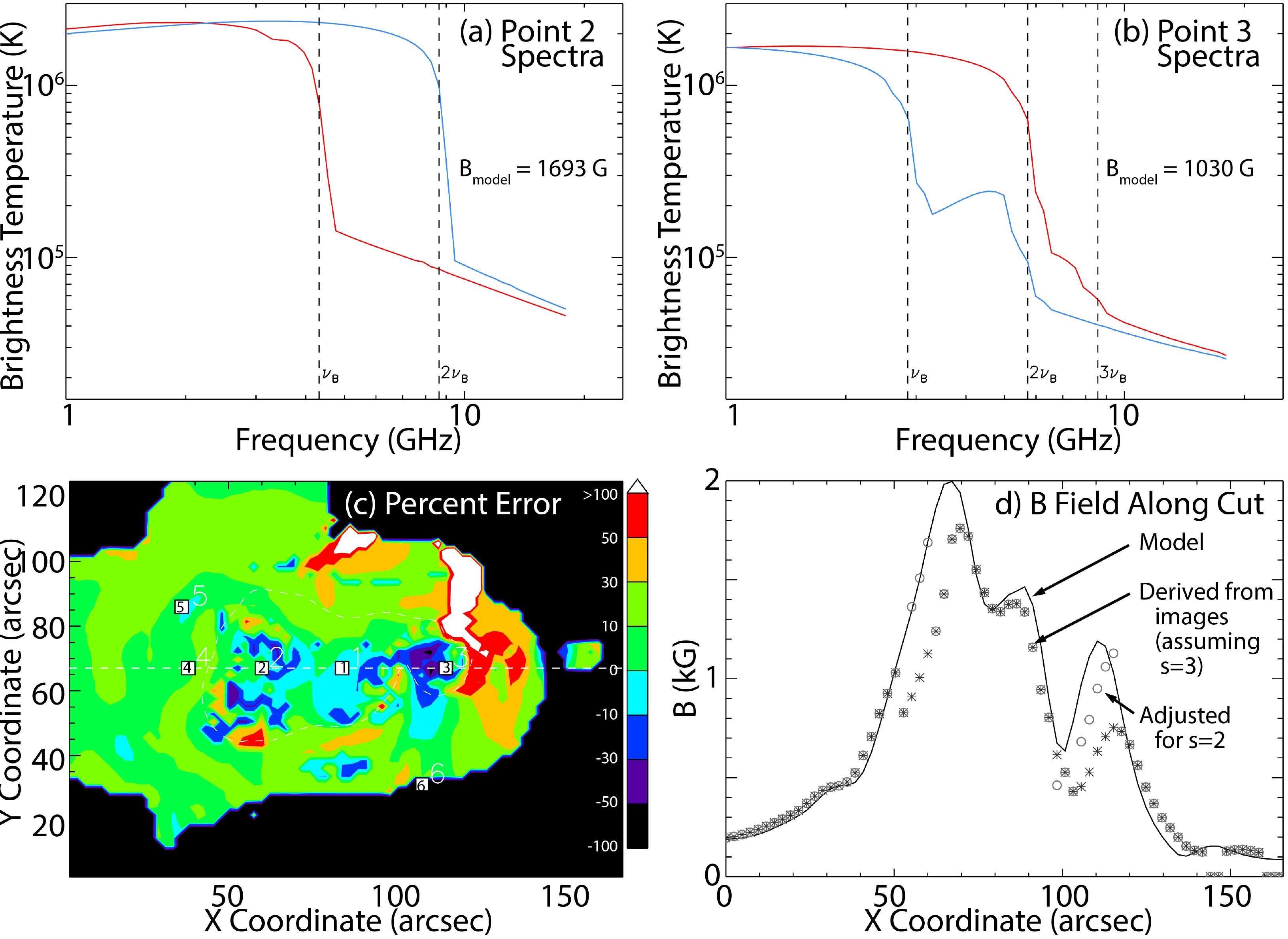}
\caption{\label{f_GR} Results obtained from simulated images derived from an active region model. (a) Example spectra from point 2 marked in (c), in right-circular polarization (RCP, red) and left-circular polarization (LCP, blue). Only two harmonics are apparent, and the magnetic field strength derived from them is 1693~G. (b) Same as (a), for point 3 marked in (c).  Note that the polarity of the field is opposite to that in panel (a).  Here again two harmonics are apparent from the spectra, with a slight indication of the third harmonic in RCP. (c) Similar spectra over the entire region have been used to derive the magnetic field strength, which was then compared with the model assuming the third harmonic ($s=3$).  Plotted is the percent error of the measurements relative to the model. (d) Asterisks show the derived magnetic field along the cut indicated by the white dashed line in (c), assuming $s=3$ compared with the model (black curve).  Points that are adjusted by a factor of 1.5 to account for $s=2$ derived from the spectra are shown as open circles.  Adapted from Wang et al. (2015).
 }
\end{figure}

Wang et al. (2015) demonstrated the technique to measure coronal magnetic fields by creating a realistic 3-dimensional model of an active region, calculating the radio images from it including both GR and FF emission, and then using the level-0 technique to derive the magnetic field strength from the images.  The derived magnetic field strength could then be compared directly with the known magnetic field strength in the model, to assess the differences, as shown in Figure~\ref{f_GR}c.  The errors were found to be within about $\pm10\%$ except in regions (blue in Figure~\ref{f_GR}c) directly over sunspots, where the opacity of the third harmonic ($s=3$) of the gyrofrequency drops due to the small angle of the magnetic field to the line of sight, and in regions in the northwestern part of the region (orange, red, and white in Figure~\ref{f_GR}) where higher density makes the FF opacity dominant. Figure~\ref{f_GR}a,b show individual spectra at two representative points in the blue areas of panel (c), where the expected spectral signature of the $s=3$ harmonic is missing or weak. Such points can be corrected based on the spectra (this is referred to as the level-1 technique in Wang et al. 2015) provided the spatial resolution is sufficiently high.  Figure~\ref{f_GR}d shows the magnetic field (along the cut indicated by the white dashed line in Figure~\ref{f_GR}c) from the model (black curve), from the measurements using the level-0 technique (black asterisks), and after adjustment using the level-1 (spectral shape) technique (open circles).  As shown by Wang et al. (2015), to use the level-1 technique requires a high spatial resolution, which will be available using the ngVLA.  In addition, the high-fidelity imaging and dynamic range of ngVLA will make it superb for exploiting such coronal magnetic field diagnostics.  Differences in magnetic field strength compared to what is expected from photospheric extrapolation methods will pinpoint the location of coronal currents, which can be quite localized within an active region.  Such currents are important clues to coronal heating and impending instability that may lead to flares and coronal mass ejections.

Whenever the GR emission is low, or the plasma density is particularly high, the dominant contribution to MW and mm emission comes from the FF process. The intensity of the FF emission is determined by the plasma density and temperature, while it only weakly depends on the magnetic field. In contrast, the polarization does depend on the magnetic field, primarily on its LOS component, which offers a way of probing this LOS component \citep{1980SoPh...67...29B,Fl_etal_2015ALMA,2017A&A...601A..43L} in addition to the absolute value available from the GR diagnostics.
This diagnostic will be available from the high-frequency ($\sim12-116$~GHz) ngVLA data at various heights of the upper chromosphere and TR. This diagnostics will  perfectly complement the GR diagnostics of the absolute value of the magnetic field. As we show in a separate paper (Fleishman et al. ApJ 2018 , under review), the probing of two components of the magnetic field, e.g., $|B|$ and $B_{los}$, is exactly what is needed to dramatically improve the coronal magnetic field reconstruction.

The brightness of the optically thick thermal emission from active regions provides a direct measurement of the plasma temperature at the height range where the given radio emission is formed. Stated another way, the radio measurements offer a unique way of quantitatively studying the magneto-thermal coupling in the corona and chromosphere. For example, the GR process can be directly used to distinguish the mechanisms of coronal heating at various heights in weak-field and strong-field regions.
This is different from the EUV diagnostics, which quantify the thermal structure (density and temperature), but yield only limited morphological information on the magnetic field.

\section{Contribution of the ngVLA}

The Sun is one of the most difficult objects to observe for a synthesis radio telescope. It has the advantage of high signal levels, but the corresponding disadvantage of needing to avoid saturation in the signal path. Further, as an imaging problem, the Sun fills the ngVLA field of view with time-varying emission on all spatial scales. An interferometer consisting of a spatially-sparse array does not sample all necessary spatial scales, and Earth-rotation synthesis cannot be used to fill in missing $uv$ spacings due to the time-variable nature of the emission.  A large number of visibility measurements is therefore needed on an appropriate distribution of spatial frequencies on short time scales; i.e., excellent instantaneous $uv$ coverage is essential. The "Plains configuration" of ngVLA antennas, comprised of 164 antennas distributed as 94 core antennas plus 74 antennas extending out to baselines comparable to the present-day JVLA A configuration, will be an excellent match to the imaging problem posed by the Sun although solar observations will mostly use the inner few km of the array, providing $\sim\,10^4$ visibilities per snapshot.

\smallskip
\smallskip

The use of 18m antennas alone would result in a lack of short-spacing information needed to measure the emission on large spatial scales, which is particularly problematic at higher frequencies where there is relatively little contrast in the image, but the addition of a short-baseline array of 6m antennas, plus total power antennas, will address this problem. Below 10 GHz, radio images of active regions generally have bright compact GR features that facilitate self-calibration, and very high dynamic range should be achievable, necessary to measure the fainter FF emission together with the bright GR sources in the same image. While the VLA and JVLA have produced many major solar results, they are limited in the effective spatial resolution that can be achieved: the longest baseline determines the number of resolution elements within the field of view, and reconstruction of the image generally requires about the same number of visibilities, but that must include short-spacing information. The VLA has no short-spacing information in its wider configurations, and just 350 baselines for dynamic snapshot imaging: ngVLA will be better in both these aspects by orders of magnitude, providing photometrically accurate imaging over the entire range of frequencies.

Another important consideration is that for frequencies $\gtrsim\,20$ GHz, where coronal scattering is no longer a significant factor, arcsecond and subarcsecond imaging will be possible. Both ground-based (including ALMA) and space-based solar observatories now routinely achieve subarcsecond imaging of the Sun: ngVLA will make it possible for solar microwave observations to match this resolution and resolve the fine-scale structure in the solar chromosphere that is a critical aspect of solar magnetic activity.

\section{Synergies with the ngVLA}
\subsection{Synergies with EUV and X-ray observations}

The characteristic thermal emissions from the solar corona occur at extreme ultraviolet (EUV) and soft X-ray (SXR) wavelengths,  which are absorbed by Earth's atmosphere and thus cannot be observed from the ground. A number of space missions, such as NASA's Solar Dynamics Observatory (SDO), provide high-spatial-resolution imaging at a number of EUV wavelengths that can be used to study the thermal properties of the corona. Such data complement the magnetic field studies that can be made with radio data, and combination of the two wavelength regimes can provide the full specification of density, temperature and magnetic field needed to describe the energetics of the Sun's atmosphere. Thus ngVLA data will be combined with EUV/SXR imaging obtained with future space missions.

\smallskip

\subsection{Synergies with mm/sub-mm observations}


At frequencies above 20 GHz, GR emission is typically not expected and solar radio emission is dominated by optically-thick thermal emission from the solar chromosphere (the few thous\-and-km thick layer immediately above the photosphere with temperatures in the range 5000$\sim$20000 K). As noted above, the chromosphere is the region of transition between the plasma-dominated photosphere (plasma thermal pressure/magnetic pressure $\beta\,>\,1$) and the magnetically-dominated corona ($\beta\,<\,1$). Convective motions driven from below make the chromosphere a very dynamic layer, changing appearance on timescales of seconds. There is a great deal of interest in how magnetic fields change with height through this layer, since that controls how much magnetic flux reaches the corona, but measurements of magnetic fields are difficult at UV wavelengths (the primary emission region for chromospheric temperatures) due at least in part to the need to integrate for some time to measure weak linear polarization. In principle radio observations, which only require Stokes I and V to determine B (line of sight), can be made more rapidly and thus can ``freeze'' the  chromospheric motion. Since the radio emission is generally optically thick above 20 GHz, there is a scaling between observing frequency and height such that higher frequencies penetrate deeper into the atmosphere. ngVLA can see most of the upper chromosphere: at 20 GHz ngVLA sees the layer with a temperature around 10000 K, while at 100 GHz emission comes from heights in the solar atmosphere where the temperature is around 7000 K. ngVLA thus complements ALMA observations at higher frequencies \citep{2017A&A...601A..43L}, which penetrate even deeper into the chromosphere (e.g., around 5900 K at 230 GHz). Measuring magnetic fields above an active region with both ngVLA and ALMA \citep{Fl_etal_2015ALMA} simultaneously offers the prospect of tracking, at high spatial resolution in all three dimensions, the behavior of magnetic field with height through a critical part of the solar atmosphere.

\subsection{Synergies with Modeling}

The problem of mapping coronal magnetic fields and tracking their evolution will likely require collaboration with both IR/UV observers measuring coronal lines as well as modelers extrapolating photospheric (or chromospheric) magnetic field measurements into the corona. Radio data complement the IR/UV techniques well, since the latter are generally confined to observations off the solar disk (i.e., just above the visible limb) due to the fact that strong emission at these wavelengths from the disk itself generally overwhelms the weaker coronal lines being used to measure polarization. Use of the Hanle effect also tends to restrict these techniques to regions of weaker field. Radio observations do not suffer from this problem and can be used anywhere on the solar disk, although they have more difficulty in measuring weak fields. Extrapolation techniques are not useful for regions near the solar limb because foreshortening due to the reduced inclination of the solar surface to the line of sight severely degrades magnetic field measurements. Thus radio observations of regions on the solar disk provide the only real ``ground-truth'' measurements of coronal magnetic fields that can be used to validate extrapolation techniques, and hence determine whether such techniques can overcome the issue of the solar photosphere not being force-free. The next step then would be adding the
%
chromospheric and coronal magnetic field diagnostics obtained from the radio (ngVLA and ALMA) data directly to the algorithms of nonlinear force-free field reconstructions of the magnetic field. As shown by Fleishman et al. (2018, under review) adding such data is capable of improving all metrics of NLFFF model by at least a factor of two; in particular, to greatly improve reconstruction of the magnetic connectivity. These improved 3D magnetic reconstructions are then needed to build realistic 3D magneto-thermal models of ARs as described by \citet{2018ApJ...853...66N}. 

%

\section{Concluding Remarks}

High-resolution multi-frequency radio data is the key to attack central problems of modern solar physics.  The extremely broad frequency coverage provided by the ngVLA in addition to its excellent instantaneous $uv$ coverage, the Plains configuration plus compact array for measuring large angular scales, a total power capability, and support of polarimetry make it both ideal and wholly unique in its ability to perform high fidelity broadband spectropolarimetry of the Sun.

The Sun's magnetic field lies at the root of a number of outstanding problems in solar physics, including magnetic energy release (flares, coronal mass ejections), coronal heating and solar wind acceleration. Quantitative measurements of coronal and chromospheric magnetic fields have been lacking to date. Such measurements can be made by broadband spectropolarimetric observations of solar active regions. These can be used synergistically with a number of other next generation instruments operating at complementary wavelength regimes, including ALMA, DKIST, and space based imagers operating in EUV and SXR wavelengths.

\acknowledgements This work was supported by NSF grants AST-1615807, AST-1735405, AST-1820613, AGS-1654382 and NASA grants NNX14AK66G, 80NSSC18K0015, NNX17AB82G, NNX14AC87G, and 80NSSC18K0667.


\bibliography{bibliography}  

\begin{thebibliography}{}
\expandafter\ifx\csname natexlab\endcsname\relax\def\natexlab#1{#1}\fi
\expandafter\ifx\csname url\endcsname\relax
  \def\url#1{\texttt{#1}}\fi
\expandafter\ifx\csname urlprefix\endcsname\relax\def\urlprefix{URL }\fi
\providecommand{\eprint}[2][]{\url{#2}}

\bibitem[{{Bogod} \& {Gelfreikh}(1980)}]{1980SoPh...67...29B}
{Bogod}, V.~M., \& {Gelfreikh}, G.~B. 1980, \solphys, 67, 29

\bibitem[{{Carlsson} et~al.(2016){Carlsson}, {Hansteen}, {Gudiksen},
  {Leenaarts}, \& {De Pontieu}}]{2016A&A...585A...4C}
{Carlsson}, M., {Hansteen}, V.~H., {Gudiksen}, B.~V., {Leenaarts}, J., \& {De
  Pontieu}, B. 2016, \aap, 585, A4. \eprint{1510.07581}

\bibitem[{{de Gouveia dal Pino} \& {Lazarian}(2005)}]{2005A&A...441..845D}
{de Gouveia dal Pino}, E.~M., \& {Lazarian}, A. 2005, \aap, 441, 845

\bibitem[{{De Rosa} et~al.(2009){De Rosa}, {Schrijver}, {Barnes}, {Leka},
  {Lites}, {Aschwanden}, {Amari}, {Canou}, {McTiernan}, {R{\'e}gnier},
  {Thalmann}, {Valori}, {Wheatland}, {Wiegelmann}, {Cheung}, {Conlon},
  {Fuhrmann}, {Inhester}, \& {Tadesse}}]{2009ApJ...696.1780D}
{De Rosa}, M.~L., {Schrijver}, C.~J., {Barnes}, G., {Leka}, K.~D., {Lites},
  B.~W., {Aschwanden}, M.~J., {Amari}, T., {Canou}, A., {McTiernan}, J.~M.,
  {R{\'e}gnier}, S., {Thalmann}, J.~K., {Valori}, G., {Wheatland}, M.~S.,
  {Wiegelmann}, T., {Cheung}, M.~C.~M., {Conlon}, P.~A., {Fuhrmann}, M.,
  {Inhester}, B., \& {Tadesse}, T. 2009, \apj, 696, 1780. \eprint{0902.1007}

\bibitem[{{De Rosa} et~al.(2015){De Rosa}, {Wheatland}, {Leka}, {Barnes},
  {Amari}, {Canou}, {Gilchrist}, {Thalmann}, {Valori}, {Wiegelmann},
  {Schrijver}, {Malanushenko}, {Sun}, \& {R{\'e}gnier}}]{2015ApJ...811..107D}
{De Rosa}, M.~L., {Wheatland}, M.~S., {Leka}, K.~D., {Barnes}, G., {Amari}, T.,
  {Canou}, A., {Gilchrist}, S.~A., {Thalmann}, J.~K., {Valori}, G.,
  {Wiegelmann}, T., {Schrijver}, C.~J., {Malanushenko}, A., {Sun}, X., \&
  {R{\'e}gnier}, S. 2015, \apj, 811, 107. \eprint{1508.05455}

\bibitem[{{Del Zanna} \& {DeLuca}(2018)}]{2018ApJ...852...52D}
{Del Zanna}, G., \& {DeLuca}, E.~E. 2018, \apj, 852, 52. \eprint{1708.03626}

\bibitem[{{Fleishman} et~al.(2015){Fleishman}, {Loukitcheva}, \&
  {Nita}}]{Fl_etal_2015ALMA}
{Fleishman}, G., {Loukitcheva}, M., \& {Nita}, G. 2015, in Revolution in
  Astronomy with ALMA: The Third Year, edited by D.~{Iono}, K.~{Tatematsu},
  A.~{Wootten}, \& L.~{Testi}, vol. 499 of Astronomical Society of the Pacific
  Conference Series, 351. \eprint{1506.08395}

\bibitem[{{Fleishman} \& {Toptygin}(2013)}]{Fl_Topt_2013_CED}
{Fleishman}, G.~D., \& {Toptygin}, I.~N. 2013, {Cosmic Electrodynamics}

\bibitem[{{Gary} \& {Hurford}(2004)}]{2004ASSL..314...71G}
{Gary}, D.~E., \& {Hurford}, G.~J. 2004, in Astrophysics and Space Science
  Library, edited by D.~E. {Gary}, \& C.~U. {Keller}, vol. 314 of Astrophysics
  and Space Science Library, 71

\bibitem[{{Gelfreikh}(2004)}]{2004ASSL..314..115G}
{Gelfreikh}, G.~B. 2004, in Astrophysics and Space Science Library, edited by
  D.~E. {Gary}, \& C.~U. {Keller}, vol. 314 of Astrophysics and Space Science
  Library, 115

\bibitem[{{Hurley} et~al.(2005){Hurley}, {Boggs}, {Smith}, {Duncan}, {Lin},
  {Zoglauer}, {Krucker}, {Hurford}, {Hudson}, {Wigger}, {Hajdas}, {Thompson},
  {Mitrofanov}, {Sanin}, {Boynton}, {Fellows}, {von Kienlin}, {Lichti}, {Rau},
  \& {Cline}}]{2005Natur.434.1098H}
{Hurley}, K., {Boggs}, S.~E., {Smith}, D.~M., {Duncan}, R.~C., {Lin}, R.,
  {Zoglauer}, A., {Krucker}, S., {Hurford}, G., {Hudson}, H., {Wigger}, C.,
  {Hajdas}, W., {Thompson}, C., {Mitrofanov}, I., {Sanin}, A., {Boynton}, W.,
  {Fellows}, C., {von Kienlin}, A., {Lichti}, G., {Rau}, A., \& {Cline}, T.
  2005, \nat, 434, 1098. \eprint{astro-ph/0502329}

\bibitem[{{Lee}(2007)}]{2007SSRv..133...73L}
{Lee}, J. 2007, \ssr, 133, 73

\bibitem[{{Loukitcheva} et~al.(2017){Loukitcheva}, {White}, {Solanki},
  {Fleishman}, \& {Carlsson}}]{2017A&A...601A..43L}
{Loukitcheva}, M., {White}, S.~M., {Solanki}, S.~K., {Fleishman}, G.~D., \&
  {Carlsson}, M. 2017, \aap, 601, A43. \eprint{1702.06018}

\bibitem[{{Nita} et~al.(2011){Nita}, {Fleishman}, {Jing}, {Lesovoi}, {Bogod},
  {Yasnov}, {Wang}, \& {Gary}}]{Nita_etal_2011}
{Nita}, G.~M., {Fleishman}, G.~D., {Jing}, J., {Lesovoi}, S.~V., {Bogod},
  V.~M., {Yasnov}, L.~V., {Wang}, H., \& {Gary}, D.~E. 2011, \apj, 737, 82.
  \eprint{1106.0262}

\bibitem[{{Nita} et~al.(2018){Nita}, {Viall}, {Klimchuk}, {Loukitcheva},
  {Gary}, {Kuznetsov}, \& {Fleishman}}]{2018ApJ...853...66N}
{Nita}, G.~M., {Viall}, N.~M., {Klimchuk}, J.~A., {Loukitcheva}, M.~A., {Gary},
  D.~E., {Kuznetsov}, A.~A., \& {Fleishman}, G.~D. 2018, \apj, 853, 66

\bibitem[{{Schopper} et~al.(1998){Schopper}, {Lesch}, \&
  {Birk}}]{1998A&A...335...26S}
{Schopper}, R., {Lesch}, H., \& {Birk}, G.~T. 1998, \aap, 335, 26.
  \eprint{astro-ph/9803329}

\bibitem[{{Sofue} et~al.(2005){Sofue}, {Kigure}, \&
  {Shibata}}]{2005PASJ...57L..39S}
{Sofue}, Y., {Kigure}, H., \& {Shibata}, K. 2005, \pasj, 57, L39.
  \eprint{astro-ph/0507568}

\bibitem[{{Tanuma} et~al.(1999){Tanuma}, {Yokoyama}, {Kudoh}, {Matsumoto},
  {Shibata}, \& {Makishima}}]{1999PASJ...51..161T}
{Tanuma}, S., {Yokoyama}, T., {Kudoh}, T., {Matsumoto}, R., {Shibata}, K., \&
  {Makishima}, K. 1999, \pasj, 51, 161

\bibitem[{{Tritschler} et~al.(2016){Tritschler}, {Rimmele}, {Berukoff},
  {Casini}, {Kuhn}, {Lin}, {Rast}, {McMullin}, {Schmidt}, {W{\"o}ger}, \&
  {DKIST Team}}]{2016AN....337.1064T}
{Tritschler}, A., {Rimmele}, T.~R., {Berukoff}, S., {Casini}, R., {Kuhn},
  J.~R., {Lin}, H., {Rast}, M.~P., {McMullin}, J.~P., {Schmidt}, W.,
  {W{\"o}ger}, F., \& {DKIST Team} 2016, Astronomische Nachrichten, 337, 1064

\bibitem[{{Tun} et~al.(2011){Tun}, {Gary}, \&
  {Georgoulis}}]{2011ApJ...728....1T}
{Tun}, S.~D., {Gary}, D.~E., \& {Georgoulis}, M.~K. 2011, \apj, 728, 1

\bibitem[{{Wang} et~al.(2015){Wang}, {Gary}, {Fleishman}, \&
  {White}}]{2015ApJ...805...93W}
{Wang}, Z., {Gary}, D.~E., {Fleishman}, G.~D., \& {White}, S.~M. 2015, \apj,
  805, 93. \eprint{1503.05239}

\end{thebibliography}



\end{document}